\documentclass[pra, aps, showpacs, notitlepage, groupedaddress,superscriptaddress, twocolumn, numerical]{revtex4-1}

\usepackage[utf8x]{inputenc}

\usepackage{amsfonts,amsmath,amssymb,stmaryrd}

\usepackage{graphicx}

\usepackage{siunitx}
\usepackage[hidelinks]{hyperref}
\usepackage[nameinlink]{cleveref}
\Crefname{equation}{Eq.}{Eqs.}
\Crefname{figure}{Fig.}{Figs.}
\Crefname{tabular}{Tab.}{Tabs.}

\usepackage{nicefrac}
\usepackage[autostyle=true]{csquotes}
\usepackage{tikz}
\usepackage{tabularx}
\usepackage{todonotes}
\usepackage{bm} 
\usepackage{color}

\newcommand{\boltzmann}{k_\mathrm{B}}

\newcommand{\ttf}{T/T_\mathrm{F}}
\newcommand{\intp}{(k_\mathrm{F} a)^{-1}}
\newcommand{\avg}[1]{\bar{#1}}
\newcommand{\epot}{E_\mathrm{pot}}

\DeclareSIUnit\gauss{G}
\DeclareSIUnit\bohrradii{a_0}

\begin{document}

	\title{Dipole oscillations of fermionic superfluids\\ along the BEC-BCS crossover in disordered potentials}
	\author{Benjamin Nagler}
	\affiliation{Department of Physics and Research Center OPTIMAS, Technische Universit\"at Kaiserslautern, Germany}
	\affiliation{Graduate School Materials Science in Mainz, Gottlieb-Daimler-Strasse 47, 67663 Kaiserslautern, Germany}	

	\author{Kevin J\"agering}
	\affiliation{Department of Physics and Research Center OPTIMAS, Technische Universit\"at Kaiserslautern, Germany}
	
	\author{Ameneh Sheikhan}
	\affiliation{Physikalisches Institut, University of Bonn, Nussallee 12, 53115 Bonn, Germany}
	
	\author{Sian Barbosa}
	\affiliation{Department of Physics and Research Center OPTIMAS, Technische Universit\"at Kaiserslautern, Germany}

	\author{Jennifer Koch}
	\affiliation{Department of Physics and Research Center OPTIMAS, Technische Universit\"at Kaiserslautern, Germany}
	
	\author{Sebastian Eggert}
	\affiliation{Department of Physics and Research Center OPTIMAS, Technische Universit\"at Kaiserslautern, Germany}

	\author{Imke Schneider}
	\affiliation{Department of Physics and Research Center OPTIMAS, Technische Universit\"at Kaiserslautern, Germany}
	
	\author{Artur Widera}
	\email{email: widera@physik.uni-kl.de}
	\affiliation{Department of Physics and Research Center OPTIMAS, Technische Universit\"at Kaiserslautern, Germany}
	\affiliation{Graduate School Materials Science in Mainz, Gottlieb-Daimler-Strasse 47, 67663 Kaiserslautern, Germany}	

	\date{\today}
	
	\begin{abstract}
		We investigate dipole oscillations of ultracold Fermi gases along the BEC-BCS crossover through disordered potentials. We observe a disorder-induced damping of oscillations as well as a change of the fundamental Kohn-mode frequency. The measurement results are compared to numerical density matrix renormalization group calculations as well as to a three-dimensional simulation of non-interacting fermions. 
		Experimentally, we find a disorder-dependent damping, which grows approximately with the second power of the disorder strength. Moreover, we observe experimentally a change of oscillation frequency which deviates from the expected behavior of a damped harmonic oscillator on a percent level. While this behavior is qualitatively expected from the theoretical models used, quantitatively the experimental observations show a significantly stronger effect than predicted by theory. Furthermore, while the frequency shift seems to scale differently with interaction strength in the BEC versus BCS regime, the damping coefficient apparently decreases with the strength of interaction, but not with the sign, which changes for BEC and BCS type Fermi gases. This is surprising, as the dominant damping mechanisms are expected to be different in the two regimes.
	
	\end{abstract}
	
	\maketitle
	
	\section{Introduction}
		Ultracold, dilute gases allow to experimentally probe fundamental properties of quantum fluids \cite{Anderson95,Davis95,OHara02}. An important question concerns the transport properties of quantum fluids in disordered media, which have been subject of intense studies since the observation of superfluid flow of helium \cite{Chan1988}. This has sparked the investigation of ultracold quantum gases in disordered media \cite{Lye05,Clement08,Dries10, Fort05}. In disordered potentials, the phenomenon of Anderson localization \cite{Anderson58}, i.e.~interference-induced absence of diffusion, has been observed in various physical realizations of quasi-noninteracting gases \cite{Billy08,Roati08,Kondov11,Jendrzejewski2012,white2019}.
		
		A fascinating feature of cold gases is the ability to additionally control the interaction strength via magnetic Feshbach resonances \cite{Inouye98}. For ultracold Fermi gases, this has opened the door to experimentally access the crossover from a molecular Bose-Einstein condensate (mBEC) via a resonantly interacting Fermi gas, to a BCS-type superfluid \cite{Regal04,Bartenstein04,Zwierlein04,Kinast04,Chin04}.
		
		The fundamental oscillation mode in a harmonic trap, the so-called Kohn mode, is not affected by interactions \cite{Kohn1961,Dobson1994}. For an additional external potential, however, the frequency and damping of the Kohn mode can sensitively indicate interactions with the environment. Dipole oscillations of a quantum gas have revealed, for example, the damping of an oscillating BEC in weak disorder \cite{Dries10} or the mutual influence of oscillating Bose- and Fermi quantum gases \cite{Delehaye15}. Moreover, it was shown that weak disorder is expected to introduce a shift of the oscillation frequency \cite{Falco07}. Furthermore, for numerical simulations of the interacting Gross-Pitaevskii equation of one-dimensional gases in disorder \cite{Hsueh18}, signatures for localization of weakly interacting gases were found in the thermalization of dipole oscillations.
		 
		Here, we study the dipole oscillation of an interacting $^6$Li Fermi gas across the BEC-BCS crossover. We focus on probing the oscillation frequency and damping of the quantum gas for different interaction scenarios along the BEC-BCS crossover. We compare our findings to one-dimensional density matrix renormalization group (DMRG) calculations as well as to a three-dimensional simulation of non-interacting fermions. 
	 
	\section{Experimental setup}
		\begin{figure}
			\includegraphics{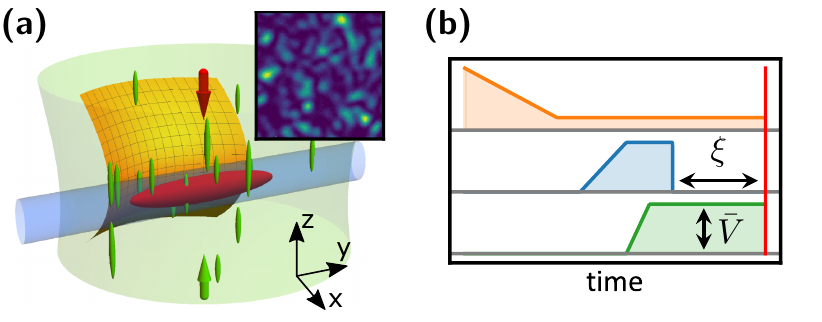}
			\caption{Schematic illustration of experimental setup and measurement sequence. (a) Experimental setup. The sample (red ellipsoid) is trapped in a superposition of an optical dipole trap (blue tube) and a magnetic saddle potential (yellow surface). The beams used for absorption imaging (red arrow) and speckle potential (green volume) propagate along the same axis in opposing directions. The inset shows a $\SI{15}{\micro\meter}\times \SI{15}{\micro\meter}$ section of the speckle intensity distribution in the $x$-$y$-plane. (b) Measurement sequence. Orange: magnetic field setting the scattering length $a$, blue: magnetic field gradient for cloud displacement, green: disorder potential strength $\avg{V}$. After a variable hold time $\xi$, the column-density distribution in the $x$-$y$-plane is recorded using absorption imaging (red line).}
			\label{fig:figure1}
		\end{figure}
	
		Experimentally, we prepare quantum gases of $N \simeq 10^6$ fermionic $^6$Li atoms by forced evaporative cooling in an equal mixture of the two lowest-lying Zeeman substates of the electronic ground state $^2\mathrm{S}_{1/2}$. Evaporation takes place in a hybrid magnetic-optical trap at a magnetic field of $\SI{840}{\gauss}$ close to a Feshbach resonance centered at \SI{832.2}{\gauss} \cite{Zuern2013}, for details of setup and sequence, see \cite{Gaenger2018}. After evaporation, the sample is held at a constant trap depth of $\SI{250}{\nano\kelvin}\times\boltzmann$ for \SI{250}{\milli\second} to ensure thermal equilibrium before the magnetic field is ramped to its final value during \SI{200}{\milli\second} (\Cref{fig:figure1}(b)), setting the required value of the $s$-wave scattering length $a$ and, thus, the interaction parameter $\intp$ \cite{Grimm2007}, with $k_\mathrm{F}$ the Fermi wave vector. The trapping frequencies are \mbox{$(\Omega_x,\Omega_y,\Omega_z)=2\pi\times(195,22.6,129)~\si{\hertz}$}, yielding the Fermi energy \mbox{$E_\mathrm{F}=\hbar(3\Omega_x \Omega_y \Omega_z N)^{1/3} \simeq \SI{600}{\nano\kelvin}$}, where $\hbar$ is the reduced Planck constant. The precise value of $\Omega_y$ depends on the magnetic field \cite{supps}.
		
		Depending on the magnitude and sign of the interaction parameter, the gas is in the BEC \mbox{($\intp\gtrsim1$)}, unitary \mbox{($\intp\approx0$)} or BCS \mbox{($\intp\lesssim-1$)} regime \cite{Grimm2007}. In the BEC regime, fermions of opposite spin form bosonic molecules.
		
		We characterize the gas in the BEC regime at a magnetic field of \SI{680}{\gauss} (\mbox{$\intp\approx 4$}), where it is possible to measure the absolute temperature by fitting the characteristic bimodal density profile \cite{Naraschewski1998}. From this, we infer the temperature \mbox{$T=\SI{150}{\nano\kelvin}$} and corresponding reduced temperature \mbox{$\ttf=\num{0.35}$}, where \mbox{$T_\mathrm{F}=E_\mathrm{F}/\boltzmann$} is the Fermi temperature and $\boltzmann$ the Boltzmann constant. According to \cite{Chen2005}, the reduced temperature $\ttf$ deep in the BEC regime is an upper bound for $\ttf$ in the strongly-interacting and BCS regime, provided that $\intp$ changes adiabatically during magnetic field ramps.
		
		The speckle potential is created by passing a laser beam of wavelength \SI{532}{\nano\meter} through a diffusive plate and focusing the light using an objective with numerical aperture \num{0.29} onto the atoms. They experience a repulsive and spatially random (but temporally constant) dipole potential $V$, which we characterize by its average $\avg{V}$ at the focal point of the objective. The typical grain size of the speckle is given by the Gaussian-shaped autocorrelation function of the potential with $1/e$ widths, i.e.~correlation lengths \cite{Kuhn2007}, \mbox{$\sigma_{x,y}=\SI{750}{\nano\meter}$} transversely to and \mbox{$\sigma_z\approx\SI{10}{\micro\meter}$} along the beam propagation direction. As the speckle beam profile has a Gaussian envelope with waist \SI{850}{\micro\meter}, the disorder potential is slightly inhomogeneous with less than \SI{10}{\percent} variation of $\avg{V}$ across the typical cloud size and oscillation trajectory.  Importantly, molecules experience double the disorder strength as they possess twice the polarizability of unbound atoms.
				
\begin{figure}
			\includegraphics{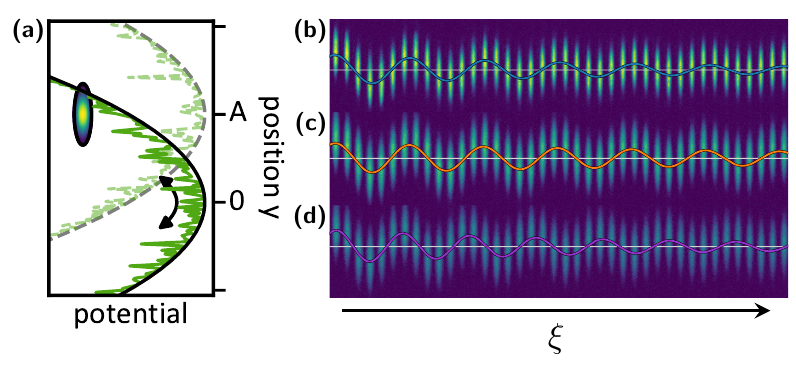}
			\caption{(a) Schematic illustration of experimental sequence. (a) and (b) and (c) Oscillation trajectories of density profiles for \mbox{$\avg{V}/\epot=\num{0.06}$} and \mbox{$\SI{0}{\milli\second}\leq\xi\leq\SI{360}{\milli\second}$}. Colored lines depict the center of mass position and white lines mark the trap center. (b) BEC regime (\mbox{$\intp=1.1$}). (c) Unitary regime (\mbox{$\intp=0$}). (d) BCS regime (\mbox{$\intp=-0.7$}).}
			\label{fig:figure2}
		\end{figure}

		In order to initiate oscillations, we displace the cloud by \mbox{$A\approx\SI{75}{\micro\meter}$} \cite{supps} along its long axis ($y$ in \Cref{fig:figure1}(a)) by application of a magnetic gradient field, which is increased during a \SI{100}{\milli\second} ramp.  
		Subsequently, the speckle disorder potential is introduced during a \SI{50}{\milli\second} linear ramp of intensity in order to minimize excitation of the gas.
		We release the cloud by suddenly extinguishing the magnetic gradient field and, therefore, almost instantaneously shifting the trap center to its initial position, see \Cref{fig:figure2}(a). The shift amplitude $A$ sets the initial potential energy \mbox{$\epot = m \Omega_y^2 A^2/2$} which drives the dipole oscillation in the combined potential of the magnetic-optical trap and the disorder. Here, $m$ is the atomic mass \mbox{$m_\mathrm{Li}\approx\SI{6}{\atomicmassunit}$} for gases in the BCS and unitary regime and the molecular mass $2m_\mathrm{Li}$ in the BEC regime.
		After variable hold times $\xi$ of up to one second, we record the atomic density distribution using resonant high-intensity absorption imaging \cite{Reinaudi2007}. The center of mass position of the cloud is extracted by fitting a 2D Thomas-Fermi profile to the measured density distribution.
		
		\Cref{fig:figure2} shows time series of cloud oscillations for disorder strength $\avg{V}/\epot=\num{0.06}$ and all three explored interaction parameters. In all cases, we observe a damped harmonic oscillation of the center of mass position.
			
	\section{Numerical DMRG calculations}
		In order to theoretically simulate the dipole oscillations as a function of time in a quantum many-body system we consider a one-dimensional (1D) version of the corresponding setup based on the Hamiltonian 
		\begin{eqnarray}
			\label{FermiGas}
			\hat{H}=& & 
			\sum_{\sigma=\uparrow,\downarrow} 
			\int \mathrm{d}x 
			\Psi_\sigma^\dagger(x) 
			\left( -\frac{\hbar^2}{2m} \partial_x^2 + \frac{1}{2} m \Omega^2 x^2 \right)
			\Psi_\sigma(x)
		\end{eqnarray}
		modeling Fermions with spin \mbox{$\sigma=-\bar\sigma$}. In the time-dependent DMRG \cite{vidal2004,daley2004,white2004} simulations we implement the model on a lattice including also disorder and interactions

		\begin{eqnarray}
			\label{FermiHubbard}
			\hat{H}= &&\sum_{i} \sum_{\sigma=\uparrow,\downarrow}  \left[ -J\left( c_{i,\sigma}^\dagger c_{i+1,\sigma}^{\phantom{\dagger} }+ \text{H.c.} \right) \right. \nonumber\\
			&& +\left.  \left(V_i + \alpha x_i
			^2\right) n_{i,\sigma} + \frac{U}{2}  n_{i,\sigma} n_{i,\overline{\sigma}}\right],
		\end{eqnarray}
		in the limit of flat traps and small densities, where the parameters are related by

		\begin{align}
			m=\frac{\hbar^2}{2Jd^2}, \phantom{abcd} \Omega  = \frac{2 d \sqrt{\alpha J}}{\hbar}, \phantom{abcd} x_i = id,
		\end{align}
		and $U$ is the on-site interactions using the local densities $n_{i,\sigma}= c_{i,\sigma}^\dagger c_{i,\sigma}^{\phantom{\dagger} }$ and lattice spacing $d$.	The uncorrelated disorder is modeled in  the form $V_i=\delta\, r_i$  with a disorder strength $\delta$ and a random value taken from the continuous uniform distribution $r_i \in [0,1]$ for each site. Specifically, we time-evolve an oscillating wave packet of $N=6$ particles with equal number of particles in each spin component in a trap with $\alpha = 0.0015~J/d^2$ after shifting by $A=8d$. The $1/e$ radius of the cloud at $U=0$ is about 10 sites. Using a time step of $\tau=\frac{1}{10}\frac{1}{J}$ we managed to resolve about $2.5$ full oscillations with disorder averaging  over 8 realizations, keeping up to $M=800$ states.
		
		It is known that for negative $U$ the 1D system crosses over from a weakly bound BCS-like state for $U=0^-$ to a BEC-like state for $U \to - \infty$ \cite{Astra04} which mimics the corresponding behavior of the 3D system.  For repulsive interactions $U>0$ the 1D system is described by dimer excitations \cite{Astra04,tokatly2004,fuchs2004}, which do not have a simple correspondence in the 3D experiment and will not be considered here. The relation between $U$ and scattering length $a$ was determined by considering a simple 1D scattering problem on a lattice to be \mbox{$a = -Jd/U$}.  The effective Fermi wave vector $k_\mathrm{F}=\sqrt{2m E_\mathrm{F}}/\hbar$ is given by the energy $E_\mathrm{F} = (N+1) \hbar \Omega/2 $ of	the highest occupied state in the trap, where we have neglected the interaction dependence \cite{soeffing11}. Therefore, we find		
		\begin{equation}
			\frac{1}{a k_\mathrm{F}} = -  \frac{U}{(N+1)^{1/2}d^{1/2} \alpha^{1/4} J^{3/4}}
		\end{equation}
		which we have used in the following to compare the simulations to experimental data.
		
	\section{Numerical simulation of non-interacting fermions}
		Additionally, we model the experiment considering non-interacting fermions in a three dimensional harmonic trap which are subjected to a speckle potential. We investigate the evolution of the atomic cloud after exciting its dipole mode by rapidly shifting the center of the trap, following the experimental sequence. In order to simulate this system, we consider the time-dependent Schr\"odinger equation $i\hbar \frac{\partial}{\partial t }|\Psi(t)\rangle = \hat{H}|\Psi(t)\rangle$ with Hamiltonian 
		\begin{eqnarray}
			\label{eq:ham}
			\hat{H}=&-&\frac{\hbar^2}{2m}\nabla^2+ \frac{1}{2}m \left(\Omega_x^2 x^2+\Omega_y^2 (y-A)^2+\Omega_z^2 z^2\right)\nonumber\\
			&+& V(x,y,z)
		\end{eqnarray} 
		where the first term is the kinetic energy of particles with mass \mbox{$m=m_\mathrm{Li}$}, the second term describes the harmonic trapping potential, with frequencies as in the experiment, centered at position $(0, A, 0)$, and the last term is the speckle potential $V$ with the spatially averaged value $\avg{V}$. In the experiment, the correlation length of the disorder $\sigma_z$ in $z$-direction is large, therefore we assume the speckle potential to be constant in this direction \mbox{($V(x,y,z)=V(x,y)$)}. We also assume that \mbox{$V(x,y)= V(x)+V(y)$}, which enables us to separate the Schr\"odinger equation in the three different dimensions. We diagonalize the discrete form of the Hamiltonian (Eq.~\ref{eq:ham}) in each direction with discrete sizes \mbox{$\Delta x = \Delta y = \Delta z$} and calculate the density of particles in time.
		
		We consider \num{2e5} fermions at zero temperature in the harmonic trapping potential shifted to position $A=\SI{71}{\micro\meter}$ along the $y$-direction. At time \mbox{$t=0$}, we quench the position of the trap to \mbox{$y=0$} and calculate the evolution of the density of atoms up to time $t=\SI{1}{\second}$ with time step \mbox{$\Delta t = \SI{2.78}{\milli\second}$}. We simulate this system for three different discrete sizes of $\Delta y = \SI{52.5}{\nano\meter}$, \SI{105}{\nano\meter} and \SI{210}{\nano\meter} where the results converge for $\Delta y \leq \SI{105}{\nano\meter}$. In the following we choose \mbox{$\Delta y = \SI{105}{\nano\meter}$} for a system of size $\SI{525}{\micro\meter} \times \SI{126}{\micro\meter} \times \SI{126}{\micro\meter}$ where the number of discrete points along different directions are $L_y= 5000$ and $L_x=L_z=1200$. We calculate the center of mass evolution in time for different disorder strengths $\avg{V}$ as considered in the experiment. For each disorder strength, the results are averaged over different realizations of disorder. When the speckle potential is off (\mbox{$\avg{V}=0$}), the center of mass shows pure dipole oscillations. 
	
	\section{Results}
		Experimentally as well as for both numerical approaches, we extract the center of mass oscillation for identical relative disorder strengths $\avg{V}/\epot$. For the DMRG calculations, also similar interaction strengths are considered. All oscillation trajectories are fitted by the classical equation of an underdamped harmonic oscillator
		\begin{equation}
			y(t) = A_\mathrm{fit} \exp\!\left(-\gamma t\right)\sin (\omega_\mathrm{fit} t + \phi) + y_0,
		\end{equation}
		with amplitude $A_\mathrm{fit}$, fitted oscillation frequency $\omega_\mathrm{fit}$, damping coefficient $\gamma$, phase $\phi$ and offset $y_0$. For a classical harmonic oscillator we expect 
		\begin{equation}
		 	\omega_\mathrm{cl} = \sqrt{\Omega_y^2 - \gamma^2},
		\end{equation} 
		and we are interested in the deviations from this expectation due to the disorder potential, as well as in the disorder-induced damping. For the experimental data, we correct for the finite curvature of the speckle envelope by introducing the corresponding trapping frequency \mbox{$\omega_\mathrm{s}\propto\sqrt{\avg{V}}$} and writing $\omega_\mathrm{cl}=\sqrt{\Omega_y^2-\gamma^2-\omega_\mathrm{s}^2}$. As \mbox{$\omega_\mathrm{s}<2\pi\times\SI{1}{\hertz}$} for all explored disorder strengths, the maximum relative change in $\omega_\mathrm{cl}$ is below $\SI{0.1}{\percent}$.
		\begin{figure}
			\includegraphics{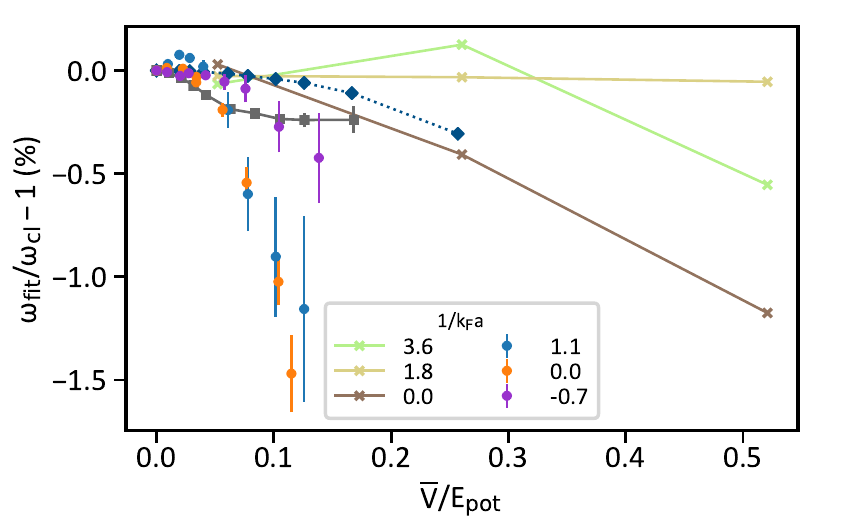}
			\caption{Relative deviation of dipole-oscillation frequency from the classical case. Dots indicate experimental data at the color-coded interaction parameter, crosses indicate results from the 1D DMRG calculation, and squares show results of the simulation of non-interacting fermions. The diamonds indicate the frequency shift predicted in \cite{Falco07}.}
			\label{fig:figure3}
		\end{figure}
		
		\Cref{fig:figure3} shows the results for the deviation of the oscillation frequency. We use the relative deviation from the expectation, i.e. $(\omega_\mathrm{fit}-\omega_\mathrm{cl}) / \omega_\mathrm{cl}$. Experimentally, we find that the frequency of oscillation shifts to smaller values at a percent level. Moreover, while the unitary gas shows the strongest shift, the oscillation frequency in the BCS regime is less affected than the BEC despite stronger interactions. 
		
\begin{figure}
			\includegraphics{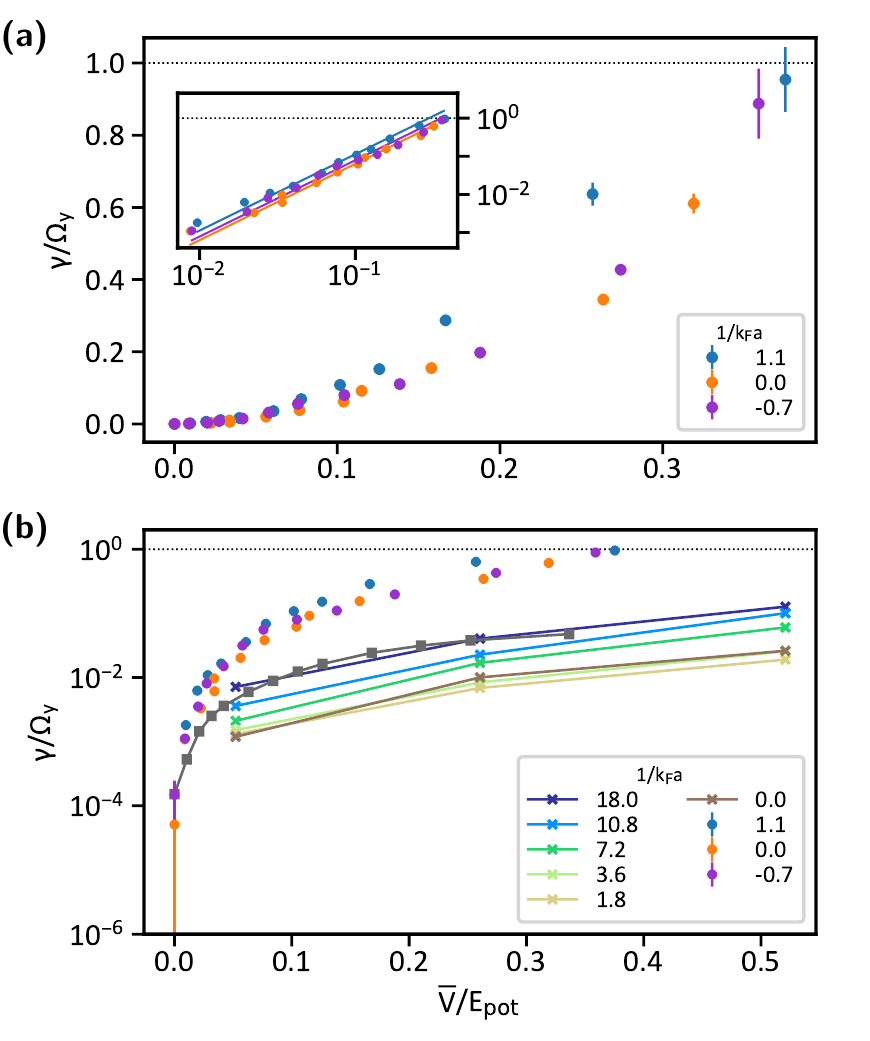}
			\caption{Damping $\gamma$ of the oscillation through disorder. Dots indicate experimental data at the color-coded interaction parameter, crosses indicate results from the 1D DMRG calculation, and squares show results of the simulation of non-interacting fermions. The dotted line marks the critical damping \mbox{$\gamma=\Omega_y$}. (a) Experimental results. The inset shows the same data in log-log scale with the solid lines being quadratic fits to the data points. (b) Comparison of experimental results to models.}
			\label{fig:figure4}
		\end{figure}

		A quantitative comparison to theoretical predictions from the models described above, and to a prediction for a BEC with healing length larger than the disorder correlation length \cite{Falco07}, shows a qualitatively similar behavior. But while the theoretical predictions all lie within the same range of oscillation shifts in the sub-percent range, the experiment shows a much stronger effect. 
		We attribute this to a combined effect of strong interactions, superfluid flow and three dimensions of the problem, which in this combination are not captured by the models we use.
		Importantly, for disorder strengths $\avg{V}/\epot \gtrsim 0.2$, the strong damping (see \Cref{fig:figure4}) obstructs the determination of an oscillation frequency from experimental data, because the system approaches the overdamped regime.
		
		The results for the damping are depicted in \Cref{fig:figure4}, where we show the fitted coefficients normalized by the disorder-free oscillation frequency $\gamma / \Omega_y$. Experimentally, we find that the damping grows approximately quadratically with disorder strength in all three interaction regimes (\Cref{fig:figure4}(a)). This is consistent with the experimental outcomes for a BEC oscillating in disorder in \cite{Dries10}. Notably, the damping is considerably weaker for the resonantly interacting superfluid and BCS gas as for the BEC, which is in accordance with the observation of superfluidity with relatively large critical velocity in the unitary regime \cite{Combescot2006,Miller2007,Weimer2015}. Moreover, it seems that the general scaling of the damping only depends on the absolute value of $\intp$ (see inset of \Cref{fig:figure4}(a)), despite the fact that dephasing and wave-like excitation of the quantum fluid is expected to prevail in the BEC, while pair-breaking might occur in the BCS regime.
		
		The theoretical models produce qualitatively the same result as the experiment, i.e. the damping increases monotonously with disorder strength. The numerical DMRG calculations show that the damping significantly increases with attractive interactions ($U<0$) towards the BEC regime, which is also seen in experiment. The magnitude of this effect, however, is in all cases more than one order of magnitude lower as in the measurements. In particular, numerical simulations show weakly damped oscillations for all disorder strengths considered, while experimentally we find the crossover to the overdamped regime for a disorder strength which corresponds to only \mbox{\textbf{$\simeq \SI{4}{\percent}$}} of the Fermi energy.
		
		We attribute this relatively strong damping to strong interactions in the cloud. At the same time, the quantitative comparison with theory shows again a much stronger experimental effect than predicted by theory. Also here, the combination of strong interactions and three spatial dimensions might explain this strongly damped dynamics.	

	\section{Conclusion}		
		The following physical picture is suggested by our investigations. For the oscillating cloud, strong interactions allow the quantum gas to react on small length scales to the disordered potential. This bending is associated with a large kinetic energy. As the quantum fluid flows, a strong energy change on small length scales will lead to excitations in the quantum gas, causing damping. At the same time, it facilitates retarding the dynamics of the gas leading to stronger reduction of the oscillation frequency with stronger interactions. The concrete quantitative description of the frequency shift and damping will be the focus of future studies, in particular the scaling with interaction parameter $\intp$ and the microscopic mechanisms underlying the transport of a BEC wave function versus the gas in the BCS regime.

		In the future it will be interesting to observe the transition from the weakly damped to the overdamped regime in order to investigate the connection between superfluid dynamics and diffusion in disorder. Moreover, the quantum phase transition to a quantum fluid is predicted to be affected by disorder \cite{Orso07, Han10}, and dipole oscillations have shown to be a sensitive tool for probing superfluid properties. 
		Furthermore, for the blue-detuned speckle potential used experimentally, the classical percolation threshold, i.e.~the energy threshold below which particles cannot explore the full potential, is around $10^{-4}\times\avg{V}$ \cite{Pilati2010}. Thus, the system is highly sensitive to reveal even the smallest fraction of localized particles if a localization transition occurs. 
		
		We thank B.~G\"anger and J.~Phieler for experimental work in the early stages of the project. We acknowledge helpful discussions with C.~Kollath and A.~Pelster. This work is funded by Deutsche Forschungsgemeinschaft (DFG, German Research Foundation) via the Collaborative Research Center SFB/TR185 (Project No. 277625399). 
		
	\bibliography{bibliography}{}

\begin{thebibliography}{46}%
\makeatletter
\providecommand \@ifxundefined [1]{%
 \@ifx{#1\undefined}
}%
\providecommand \@ifnum [1]{%
 \ifnum #1\expandafter \@firstoftwo
 \else \expandafter \@secondoftwo
 \fi
}%
\providecommand \@ifx [1]{%
 \ifx #1\expandafter \@firstoftwo
 \else \expandafter \@secondoftwo
 \fi
}%
\providecommand \natexlab [1]{#1}%
\providecommand \enquote  [1]{``#1''}%
\providecommand \bibnamefont  [1]{#1}%
\providecommand \bibfnamefont [1]{#1}%
\providecommand \citenamefont [1]{#1}%
\providecommand \href@noop [0]{\@secondoftwo}%
\providecommand \href [0]{\begingroup \@sanitize@url \@href}%
\providecommand \@href[1]{\@@startlink{#1}\@@href}%
\providecommand \@@href[1]{\endgroup#1\@@endlink}%
\providecommand \@sanitize@url [0]{\catcode `\\12\catcode `\$12\catcode
  `\&12\catcode `\#12\catcode `\^12\catcode `\_12\catcode `\%12\relax}%
\providecommand \@@startlink[1]{}%
\providecommand \@@endlink[0]{}%
\providecommand \url  [0]{\begingroup\@sanitize@url \@url }%
\providecommand \@url [1]{\endgroup\@href {#1}{\urlprefix }}%
\providecommand \urlprefix  [0]{URL }%
\providecommand \Eprint [0]{\href }%
\providecommand \doibase [0]{http://dx.doi.org/}%
\providecommand \selectlanguage [0]{\@gobble}%
\providecommand \bibinfo  [0]{\@secondoftwo}%
\providecommand \bibfield  [0]{\@secondoftwo}%
\providecommand \translation [1]{[#1]}%
\providecommand \BibitemOpen [0]{}%
\providecommand \bibitemStop [0]{}%
\providecommand \bibitemNoStop [0]{.\EOS\space}%
\providecommand \EOS [0]{\spacefactor3000\relax}%
\providecommand \BibitemShut  [1]{\csname bibitem#1\endcsname}%
\let\auto@bib@innerbib\@empty
\bibitem [{\citenamefont {Anderson}\ \emph {et~al.}(1995)\citenamefont
  {Anderson}, \citenamefont {Ensher}, \citenamefont {Matthews}, \citenamefont
  {Wieman},\ and\ \citenamefont {Cornell}}]{Anderson95}%
  \BibitemOpen
  \bibfield  {author} {\bibinfo {author} {\bibfnamefont {M.~H.}\ \bibnamefont
  {Anderson}}, \bibinfo {author} {\bibfnamefont {J.~R.}\ \bibnamefont
  {Ensher}}, \bibinfo {author} {\bibfnamefont {M.~R.}\ \bibnamefont
  {Matthews}}, \bibinfo {author} {\bibfnamefont {C.~E.}\ \bibnamefont
  {Wieman}}, \ and\ \bibinfo {author} {\bibfnamefont {E.~A.}\ \bibnamefont
  {Cornell}},\ }\href@noop {} {\bibfield  {journal} {\bibinfo  {journal}
  {Science}\ }\textbf {\bibinfo {volume} {269}},\ \bibinfo {pages} {198}
  (\bibinfo {year} {1995})}\BibitemShut {NoStop}%
\bibitem [{\citenamefont {Davis}\ \emph {et~al.}(1995)\citenamefont {Davis},
  \citenamefont {Mewes}, \citenamefont {Andrews}, \citenamefont {van Druten},
  \citenamefont {Durfee}, \citenamefont {Kurn},\ and\ \citenamefont
  {Ketterle}}]{Davis95}%
  \BibitemOpen
  \bibfield  {author} {\bibinfo {author} {\bibfnamefont {K.~B.}\ \bibnamefont
  {Davis}}, \bibinfo {author} {\bibfnamefont {M.~O.}\ \bibnamefont {Mewes}},
  \bibinfo {author} {\bibfnamefont {M.~R.}\ \bibnamefont {Andrews}}, \bibinfo
  {author} {\bibfnamefont {N.~J.}\ \bibnamefont {van Druten}}, \bibinfo
  {author} {\bibfnamefont {D.~S.}\ \bibnamefont {Durfee}}, \bibinfo {author}
  {\bibfnamefont {D.~M.}\ \bibnamefont {Kurn}}, \ and\ \bibinfo {author}
  {\bibfnamefont {W.}~\bibnamefont {Ketterle}},\ }\href {\doibase
  10.1103/PhysRevLett.75.3969} {\bibfield  {journal} {\bibinfo  {journal}
  {Phys. Rev. Lett.}\ }\textbf {\bibinfo {volume} {75}},\ \bibinfo {pages}
  {3969} (\bibinfo {year} {1995})}\BibitemShut {NoStop}%
\bibitem [{\citenamefont {O'Hara}\ \emph {et~al.}(2002)\citenamefont {O'Hara},
  \citenamefont {Hemmer}, \citenamefont {Gehm}, \citenamefont {Granade},\ and\
  \citenamefont {Thomas}}]{OHara02}%
  \BibitemOpen
  \bibfield  {author} {\bibinfo {author} {\bibfnamefont {K.~M.}\ \bibnamefont
  {O'Hara}}, \bibinfo {author} {\bibfnamefont {S.~L.}\ \bibnamefont {Hemmer}},
  \bibinfo {author} {\bibfnamefont {M.~E.}\ \bibnamefont {Gehm}}, \bibinfo
  {author} {\bibfnamefont {S.~R.}\ \bibnamefont {Granade}}, \ and\ \bibinfo
  {author} {\bibfnamefont {J.~E.}\ \bibnamefont {Thomas}},\ }\href {\doibase
  10.1126/science.1079107} {\bibfield  {journal} {\bibinfo  {journal}
  {Science}\ }\textbf {\bibinfo {volume} {298}},\ \bibinfo {pages} {2179}
  (\bibinfo {year} {2002})}\BibitemShut {NoStop}%
\bibitem [{\citenamefont {Chan}\ \emph {et~al.}(1988)\citenamefont {Chan},
  \citenamefont {Blum}, \citenamefont {Murphy}, \citenamefont {Wong},\ and\
  \citenamefont {Reppy}}]{Chan1988}%
  \BibitemOpen
  \bibfield  {author} {\bibinfo {author} {\bibfnamefont {M.~H.~W.}\
  \bibnamefont {Chan}}, \bibinfo {author} {\bibfnamefont {K.~I.}\ \bibnamefont
  {Blum}}, \bibinfo {author} {\bibfnamefont {S.~Q.}\ \bibnamefont {Murphy}},
  \bibinfo {author} {\bibfnamefont {G.~K.~S.}\ \bibnamefont {Wong}}, \ and\
  \bibinfo {author} {\bibfnamefont {J.~D.}\ \bibnamefont {Reppy}},\ }\href
  {\doibase 10.1103/PhysRevLett.61.1950} {\bibfield  {journal} {\bibinfo
  {journal} {Phys. Rev. Lett.}\ }\textbf {\bibinfo {volume} {61}},\ \bibinfo
  {pages} {1950} (\bibinfo {year} {1988})}\BibitemShut {NoStop}%
\bibitem [{\citenamefont {Lye}\ \emph {et~al.}(2005)\citenamefont {Lye},
  \citenamefont {Fallani}, \citenamefont {Modugno}, \citenamefont {Wiersma},
  \citenamefont {Fort},\ and\ \citenamefont {Inguscio}}]{Lye05}%
  \BibitemOpen
  \bibfield  {author} {\bibinfo {author} {\bibfnamefont {J.~E.}\ \bibnamefont
  {Lye}}, \bibinfo {author} {\bibfnamefont {L.}~\bibnamefont {Fallani}},
  \bibinfo {author} {\bibfnamefont {M.}~\bibnamefont {Modugno}}, \bibinfo
  {author} {\bibfnamefont {D.~S.}\ \bibnamefont {Wiersma}}, \bibinfo {author}
  {\bibfnamefont {C.}~\bibnamefont {Fort}}, \ and\ \bibinfo {author}
  {\bibfnamefont {M.}~\bibnamefont {Inguscio}},\ }\href {\doibase
  10.1103/PhysRevLett.95.070401} {\bibfield  {journal} {\bibinfo  {journal}
  {Phys. Rev. Lett.}\ }\textbf {\bibinfo {volume} {95}},\ \bibinfo {pages}
  {070401} (\bibinfo {year} {2005})}\BibitemShut {NoStop}%
\bibitem [{\citenamefont {Cl\'ement}\ \emph {et~al.}(2008)\citenamefont
  {Cl\'ement}, \citenamefont {Bouyer}, \citenamefont {Aspect},\ and\
  \citenamefont {Sanchez-Palencia}}]{Clement08}%
  \BibitemOpen
  \bibfield  {author} {\bibinfo {author} {\bibfnamefont {D.}~\bibnamefont
  {Cl\'ement}}, \bibinfo {author} {\bibfnamefont {P.}~\bibnamefont {Bouyer}},
  \bibinfo {author} {\bibfnamefont {A.}~\bibnamefont {Aspect}}, \ and\ \bibinfo
  {author} {\bibfnamefont {L.}~\bibnamefont {Sanchez-Palencia}},\ }\href
  {\doibase 10.1103/PhysRevA.77.033631} {\bibfield  {journal} {\bibinfo
  {journal} {Phys. Rev. A}\ }\textbf {\bibinfo {volume} {77}},\ \bibinfo
  {pages} {033631} (\bibinfo {year} {2008})}\BibitemShut {NoStop}%
\bibitem [{\citenamefont {Dries}\ \emph {et~al.}(2010)\citenamefont {Dries},
  \citenamefont {Pollack}, \citenamefont {Hitchcock},\ and\ \citenamefont
  {Hulet}}]{Dries10}%
  \BibitemOpen
  \bibfield  {author} {\bibinfo {author} {\bibfnamefont {D.}~\bibnamefont
  {Dries}}, \bibinfo {author} {\bibfnamefont {S.~E.}\ \bibnamefont {Pollack}},
  \bibinfo {author} {\bibfnamefont {J.~M.}\ \bibnamefont {Hitchcock}}, \ and\
  \bibinfo {author} {\bibfnamefont {R.~G.}\ \bibnamefont {Hulet}},\ }\href
  {\doibase 10.1103/PhysRevA.82.033603} {\bibfield  {journal} {\bibinfo
  {journal} {Phys. Rev. A}\ }\textbf {\bibinfo {volume} {82}},\ \bibinfo
  {pages} {033603} (\bibinfo {year} {2010})}\BibitemShut {NoStop}%
\bibitem [{\citenamefont {Fort}\ \emph {et~al.}(2005)\citenamefont {Fort},
  \citenamefont {Fallani}, \citenamefont {Guarrera}, \citenamefont {Lye},
  \citenamefont {Modugno}, \citenamefont {Wiersma},\ and\ \citenamefont
  {Inguscio}}]{Fort05}%
  \BibitemOpen
  \bibfield  {author} {\bibinfo {author} {\bibfnamefont {C.}~\bibnamefont
  {Fort}}, \bibinfo {author} {\bibfnamefont {L.}~\bibnamefont {Fallani}},
  \bibinfo {author} {\bibfnamefont {V.}~\bibnamefont {Guarrera}}, \bibinfo
  {author} {\bibfnamefont {J.~E.}\ \bibnamefont {Lye}}, \bibinfo {author}
  {\bibfnamefont {M.}~\bibnamefont {Modugno}}, \bibinfo {author} {\bibfnamefont
  {D.~S.}\ \bibnamefont {Wiersma}}, \ and\ \bibinfo {author} {\bibfnamefont
  {M.}~\bibnamefont {Inguscio}},\ }\href {\doibase
  10.1103/PhysRevLett.95.170410} {\bibfield  {journal} {\bibinfo  {journal}
  {Phys. Rev. Lett.}\ }\textbf {\bibinfo {volume} {95}},\ \bibinfo {pages}
  {170410} (\bibinfo {year} {2005})}\BibitemShut {NoStop}%
\bibitem [{\citenamefont {Anderson}(1958)}]{Anderson58}%
  \BibitemOpen
  \bibfield  {author} {\bibinfo {author} {\bibfnamefont {P.~W.}\ \bibnamefont
  {Anderson}},\ }\href {\doibase 10.1103/PhysRev.109.1492} {\bibfield
  {journal} {\bibinfo  {journal} {Phys. Rev.}\ }\textbf {\bibinfo {volume}
  {109}},\ \bibinfo {pages} {1492} (\bibinfo {year} {1958})}\BibitemShut
  {NoStop}%
\bibitem [{\citenamefont {Billy}\ \emph {et~al.}(2008)\citenamefont {Billy},
  \citenamefont {Josse}, \citenamefont {Zuo}, \citenamefont {Bernard},
  \citenamefont {Hambrecht}, \citenamefont {Lugan}, \citenamefont
  {Cl\'{e}ment}, \citenamefont {Sanchez-Palencia}, \citenamefont {Bouyer},\
  and\ \citenamefont {Aspect}}]{Billy08}%
  \BibitemOpen
  \bibfield  {author} {\bibinfo {author} {\bibfnamefont {J.}~\bibnamefont
  {Billy}}, \bibinfo {author} {\bibfnamefont {V.}~\bibnamefont {Josse}},
  \bibinfo {author} {\bibfnamefont {Z.}~\bibnamefont {Zuo}}, \bibinfo {author}
  {\bibfnamefont {A.}~\bibnamefont {Bernard}}, \bibinfo {author} {\bibfnamefont
  {B.}~\bibnamefont {Hambrecht}}, \bibinfo {author} {\bibfnamefont
  {P.}~\bibnamefont {Lugan}}, \bibinfo {author} {\bibfnamefont
  {D.}~\bibnamefont {Cl\'{e}ment}}, \bibinfo {author} {\bibfnamefont
  {L.}~\bibnamefont {Sanchez-Palencia}}, \bibinfo {author} {\bibfnamefont
  {P.}~\bibnamefont {Bouyer}}, \ and\ \bibinfo {author} {\bibfnamefont
  {A.}~\bibnamefont {Aspect}},\ }\href {\doibase 10.1038/nature07000}
  {\bibfield  {journal} {\bibinfo  {journal} {Nature}\ }\textbf {\bibinfo
  {volume} {453}},\ \bibinfo {pages} {891} (\bibinfo {year}
  {2008})}\BibitemShut {NoStop}%
\bibitem [{\citenamefont {Roati}\ \emph {et~al.}(2008)\citenamefont {Roati},
  \citenamefont {D'Errico}, \citenamefont {Fallani}, \citenamefont {Fattori},
  \citenamefont {Fort}, \citenamefont {Zaccanti}, \citenamefont {Modugno},\
  and\ \citenamefont {Inguscio}}]{Roati08}%
  \BibitemOpen
  \bibfield  {author} {\bibinfo {author} {\bibfnamefont {G.}~\bibnamefont
  {Roati}}, \bibinfo {author} {\bibfnamefont {C.}~\bibnamefont {D'Errico}},
  \bibinfo {author} {\bibfnamefont {L.}~\bibnamefont {Fallani}}, \bibinfo
  {author} {\bibfnamefont {M.}~\bibnamefont {Fattori}}, \bibinfo {author}
  {\bibfnamefont {C.}~\bibnamefont {Fort}}, \bibinfo {author} {\bibfnamefont
  {M.}~\bibnamefont {Zaccanti}}, \bibinfo {author} {\bibfnamefont
  {G.}~\bibnamefont {Modugno}}, \ and\ \bibinfo {author} {\bibfnamefont {M.~M.
  .~M.}\ \bibnamefont {Inguscio}},\ }\href {\doibase 10.1038/nature07071}
  {\bibfield  {journal} {\bibinfo  {journal} {Nature}\ }\textbf {\bibinfo
  {volume} {453}},\ \bibinfo {pages} {895} (\bibinfo {year}
  {2008})}\BibitemShut {NoStop}%
\bibitem [{\citenamefont {Kondov}\ \emph {et~al.}(2011)\citenamefont {Kondov},
  \citenamefont {McGehee}, \citenamefont {Zirbel}, ,\ and\ \citenamefont
  {DeMarco}}]{Kondov11}%
  \BibitemOpen
  \bibfield  {author} {\bibinfo {author} {\bibfnamefont {S.}~\bibnamefont
  {Kondov}}, \bibinfo {author} {\bibfnamefont {W.}~\bibnamefont {McGehee}},
  \bibinfo {author} {\bibfnamefont {J.}~\bibnamefont {Zirbel}}, , \ and\
  \bibinfo {author} {\bibfnamefont {B.}~\bibnamefont {DeMarco}},\ }\href
  {\doibase 10.1126/science.1209019} {\bibfield  {journal} {\bibinfo  {journal}
  {Science}\ }\textbf {\bibinfo {volume} {334}},\ \bibinfo {pages} {66}
  (\bibinfo {year} {2011})}\BibitemShut {NoStop}%
\bibitem [{\citenamefont {{Jendrzejewski}}\ \emph {et~al.}(2012)\citenamefont
  {{Jendrzejewski}}, \citenamefont {{Bernard}}, \citenamefont {{M{\"u}ller}},
  \citenamefont {{Cheinet}}, \citenamefont {{Josse}}, \citenamefont {{Piraud}},
  \citenamefont {{Pezz{\'e}}}, \citenamefont {{Sanchez-Palencia}},
  \citenamefont {{Aspect}},\ and\ \citenamefont
  {{Bouyer}}}]{Jendrzejewski2012}%
  \BibitemOpen
  \bibfield  {author} {\bibinfo {author} {\bibfnamefont {F.}~\bibnamefont
  {{Jendrzejewski}}}, \bibinfo {author} {\bibfnamefont {A.}~\bibnamefont
  {{Bernard}}}, \bibinfo {author} {\bibfnamefont {K.}~\bibnamefont
  {{M{\"u}ller}}}, \bibinfo {author} {\bibfnamefont {P.}~\bibnamefont
  {{Cheinet}}}, \bibinfo {author} {\bibfnamefont {V.}~\bibnamefont {{Josse}}},
  \bibinfo {author} {\bibfnamefont {M.}~\bibnamefont {{Piraud}}}, \bibinfo
  {author} {\bibfnamefont {L.}~\bibnamefont {{Pezz{\'e}}}}, \bibinfo {author}
  {\bibfnamefont {L.}~\bibnamefont {{Sanchez-Palencia}}}, \bibinfo {author}
  {\bibfnamefont {A.}~\bibnamefont {{Aspect}}}, \ and\ \bibinfo {author}
  {\bibfnamefont {P.}~\bibnamefont {{Bouyer}}},\ }\href {\doibase
  10.1038/nphys2256} {\bibfield  {journal} {\bibinfo  {journal} {Nature
  Physics}\ }\textbf {\bibinfo {volume} {8}},\ \bibinfo {pages} {398} (\bibinfo
  {year} {2012})}\BibitemShut {NoStop}%
\bibitem [{\citenamefont {White}\ \emph {et~al.}(2019)\citenamefont {White},
  \citenamefont {Haase}, \citenamefont {Brown}, \citenamefont {Hoogerland},
  \citenamefont {Najafabadi}, \citenamefont {Helm}, \citenamefont {Gies},
  \citenamefont {Schumayer},\ and\ \citenamefont {Hutchinson}}]{white2019}%
  \BibitemOpen
  \bibfield  {author} {\bibinfo {author} {\bibfnamefont {D.~H.}\ \bibnamefont
  {White}}, \bibinfo {author} {\bibfnamefont {T.~A.}\ \bibnamefont {Haase}},
  \bibinfo {author} {\bibfnamefont {D.~J.}\ \bibnamefont {Brown}}, \bibinfo
  {author} {\bibfnamefont {M.~D.}\ \bibnamefont {Hoogerland}}, \bibinfo
  {author} {\bibfnamefont {M.~S.}\ \bibnamefont {Najafabadi}}, \bibinfo
  {author} {\bibfnamefont {J.~L.}\ \bibnamefont {Helm}}, \bibinfo {author}
  {\bibfnamefont {C.}~\bibnamefont {Gies}}, \bibinfo {author} {\bibfnamefont
  {D.}~\bibnamefont {Schumayer}}, \ and\ \bibinfo {author} {\bibfnamefont
  {D.~A.~W.}\ \bibnamefont {Hutchinson}},\ }\href@noop {} {} (\bibinfo {year}
  {2019}),\ \Eprint {http://arxiv.org/abs/1911.04858} {arXiv:1911.04858}
  \BibitemShut {NoStop}%
\bibitem [{\citenamefont {Inouye}\ \emph {et~al.}(1998)\citenamefont {Inouye},
  \citenamefont {Andrews}, \citenamefont {Stenger}, \citenamefont {Miesner},
  \citenamefont {Stamper-Kurn},\ and\ \citenamefont {Ketterle}}]{Inouye98}%
  \BibitemOpen
  \bibfield  {author} {\bibinfo {author} {\bibfnamefont {S.}~\bibnamefont
  {Inouye}}, \bibinfo {author} {\bibfnamefont {M.~R.}\ \bibnamefont {Andrews}},
  \bibinfo {author} {\bibfnamefont {J.}~\bibnamefont {Stenger}}, \bibinfo
  {author} {\bibfnamefont {H.~J.}\ \bibnamefont {Miesner}}, \bibinfo {author}
  {\bibfnamefont {D.~M.}\ \bibnamefont {Stamper-Kurn}}, \ and\ \bibinfo
  {author} {\bibfnamefont {W.}~\bibnamefont {Ketterle}},\ }\href {\doibase
  doi:10.1038/32354} {\bibfield  {journal} {\bibinfo  {journal} {Nature}\
  }\textbf {\bibinfo {volume} {392}},\ \bibinfo {pages} {151} (\bibinfo {year}
  {1998})}\BibitemShut {NoStop}%
\bibitem [{\citenamefont {Regal}\ \emph {et~al.}(2004)\citenamefont {Regal},
  \citenamefont {Greiner},\ and\ \citenamefont {Jin}}]{Regal04}%
  \BibitemOpen
  \bibfield  {author} {\bibinfo {author} {\bibfnamefont {C.~A.}\ \bibnamefont
  {Regal}}, \bibinfo {author} {\bibfnamefont {M.}~\bibnamefont {Greiner}}, \
  and\ \bibinfo {author} {\bibfnamefont {D.~S.}\ \bibnamefont {Jin}},\ }\href
  {\doibase 10.1103/PhysRevLett.92.040403} {\bibfield  {journal} {\bibinfo
  {journal} {Phys. Rev. Lett.}\ }\textbf {\bibinfo {volume} {92}},\ \bibinfo
  {pages} {040403} (\bibinfo {year} {2004})}\BibitemShut {NoStop}%
\bibitem [{\citenamefont {Bartenstein}\ \emph {et~al.}(2004)\citenamefont
  {Bartenstein}, \citenamefont {Altmeyer}, \citenamefont {Riedl}, \citenamefont
  {Jochim}, \citenamefont {Chin}, \citenamefont {Denschlag},\ and\
  \citenamefont {Grimm}}]{Bartenstein04}%
  \BibitemOpen
  \bibfield  {author} {\bibinfo {author} {\bibfnamefont {M.}~\bibnamefont
  {Bartenstein}}, \bibinfo {author} {\bibfnamefont {A.}~\bibnamefont
  {Altmeyer}}, \bibinfo {author} {\bibfnamefont {S.}~\bibnamefont {Riedl}},
  \bibinfo {author} {\bibfnamefont {S.}~\bibnamefont {Jochim}}, \bibinfo
  {author} {\bibfnamefont {C.}~\bibnamefont {Chin}}, \bibinfo {author}
  {\bibfnamefont {J.~H.}\ \bibnamefont {Denschlag}}, \ and\ \bibinfo {author}
  {\bibfnamefont {R.}~\bibnamefont {Grimm}},\ }\href {\doibase
  10.1103/PhysRevLett.92.120401} {\bibfield  {journal} {\bibinfo  {journal}
  {Phys. Rev. Lett.}\ }\textbf {\bibinfo {volume} {92}},\ \bibinfo {pages}
  {120401} (\bibinfo {year} {2004})}\BibitemShut {NoStop}%
\bibitem [{\citenamefont {Zwierlein}\ \emph {et~al.}(2004)\citenamefont
  {Zwierlein}, \citenamefont {Stan}, \citenamefont {Schunck}, \citenamefont
  {Raupach}, \citenamefont {Kerman},\ and\ \citenamefont
  {Ketterle}}]{Zwierlein04}%
  \BibitemOpen
  \bibfield  {author} {\bibinfo {author} {\bibfnamefont {M.~W.}\ \bibnamefont
  {Zwierlein}}, \bibinfo {author} {\bibfnamefont {C.~A.}\ \bibnamefont {Stan}},
  \bibinfo {author} {\bibfnamefont {C.~H.}\ \bibnamefont {Schunck}}, \bibinfo
  {author} {\bibfnamefont {S.~M.~F.}\ \bibnamefont {Raupach}}, \bibinfo
  {author} {\bibfnamefont {A.~J.}\ \bibnamefont {Kerman}}, \ and\ \bibinfo
  {author} {\bibfnamefont {W.}~\bibnamefont {Ketterle}},\ }\href {\doibase
  10.1103/PhysRevLett.92.120403} {\bibfield  {journal} {\bibinfo  {journal}
  {Phys. Rev. Lett.}\ }\textbf {\bibinfo {volume} {92}},\ \bibinfo {pages}
  {120403} (\bibinfo {year} {2004})}\BibitemShut {NoStop}%
\bibitem [{\citenamefont {Kinast}\ \emph {et~al.}(2004)\citenamefont {Kinast},
  \citenamefont {Hemmer}, \citenamefont {Gehm}, \citenamefont {Turlapov},\ and\
  \citenamefont {Thomas}}]{Kinast04}%
  \BibitemOpen
  \bibfield  {author} {\bibinfo {author} {\bibfnamefont {J.}~\bibnamefont
  {Kinast}}, \bibinfo {author} {\bibfnamefont {S.~L.}\ \bibnamefont {Hemmer}},
  \bibinfo {author} {\bibfnamefont {M.~E.}\ \bibnamefont {Gehm}}, \bibinfo
  {author} {\bibfnamefont {A.}~\bibnamefont {Turlapov}}, \ and\ \bibinfo
  {author} {\bibfnamefont {J.~E.}\ \bibnamefont {Thomas}},\ }\href {\doibase
  10.1103/PhysRevLett.92.150402} {\bibfield  {journal} {\bibinfo  {journal}
  {Phys. Rev. Lett.}\ }\textbf {\bibinfo {volume} {92}},\ \bibinfo {pages}
  {150402} (\bibinfo {year} {2004})}\BibitemShut {NoStop}%
\bibitem [{\citenamefont {Chin}\ \emph {et~al.}(2004)\citenamefont {Chin},
  \citenamefont {Bartenstein}, \citenamefont {Altmeyer}, \citenamefont {Riedl},
  \citenamefont {Jochim}, \citenamefont {Denschlag},\ and\ \citenamefont
  {Grimm}}]{Chin04}%
  \BibitemOpen
  \bibfield  {author} {\bibinfo {author} {\bibfnamefont {C.}~\bibnamefont
  {Chin}}, \bibinfo {author} {\bibfnamefont {M.}~\bibnamefont {Bartenstein}},
  \bibinfo {author} {\bibfnamefont {A.}~\bibnamefont {Altmeyer}}, \bibinfo
  {author} {\bibfnamefont {S.}~\bibnamefont {Riedl}}, \bibinfo {author}
  {\bibfnamefont {S.}~\bibnamefont {Jochim}}, \bibinfo {author} {\bibfnamefont
  {J.~H.}\ \bibnamefont {Denschlag}}, \ and\ \bibinfo {author} {\bibfnamefont
  {R.}~\bibnamefont {Grimm}},\ }\href {\doibase 10.1126/science.1100818}
  {\bibfield  {journal} {\bibinfo  {journal} {Science}\ }\textbf {\bibinfo
  {volume} {305}},\ \bibinfo {pages} {1128} (\bibinfo {year}
  {2004})}\BibitemShut {NoStop}%
\bibitem [{\citenamefont {Kohn}(1961)}]{Kohn1961}%
  \BibitemOpen
  \bibfield  {author} {\bibinfo {author} {\bibfnamefont {W.}~\bibnamefont
  {Kohn}},\ }\href {\doibase 10.1103/PhysRev.123.1242} {\bibfield  {journal}
  {\bibinfo  {journal} {Phys. Rev.}\ }\textbf {\bibinfo {volume} {123}},\
  \bibinfo {pages} {1242} (\bibinfo {year} {1961})}\BibitemShut {NoStop}%
\bibitem [{\citenamefont {Dobson}(1994)}]{Dobson1994}%
  \BibitemOpen
  \bibfield  {author} {\bibinfo {author} {\bibfnamefont {J.~F.}\ \bibnamefont
  {Dobson}},\ }\href {\doibase 10.1103/PhysRevLett.73.2244} {\bibfield
  {journal} {\bibinfo  {journal} {Phys. Rev. Lett.}\ }\textbf {\bibinfo
  {volume} {73}},\ \bibinfo {pages} {2244} (\bibinfo {year}
  {1994})}\BibitemShut {NoStop}%
\bibitem [{\citenamefont {Delehaye}\ \emph {et~al.}(2015)\citenamefont
  {Delehaye}, \citenamefont {Laurent}, \citenamefont {Ferrier-Barbut},
  \citenamefont {Jin}, \citenamefont {Chevy},\ and\ \citenamefont
  {Salomon}}]{Delehaye15}%
  \BibitemOpen
  \bibfield  {author} {\bibinfo {author} {\bibfnamefont {M.}~\bibnamefont
  {Delehaye}}, \bibinfo {author} {\bibfnamefont {S.}~\bibnamefont {Laurent}},
  \bibinfo {author} {\bibfnamefont {I.}~\bibnamefont {Ferrier-Barbut}},
  \bibinfo {author} {\bibfnamefont {S.}~\bibnamefont {Jin}}, \bibinfo {author}
  {\bibfnamefont {F.}~\bibnamefont {Chevy}}, \ and\ \bibinfo {author}
  {\bibfnamefont {C.}~\bibnamefont {Salomon}},\ }\href {\doibase
  10.1103/PhysRevLett.115.265303} {\bibfield  {journal} {\bibinfo  {journal}
  {Phys. Rev. Lett.}\ }\textbf {\bibinfo {volume} {115}},\ \bibinfo {pages}
  {265303} (\bibinfo {year} {2015})}\BibitemShut {NoStop}%
\bibitem [{\citenamefont {Falco}\ \emph {et~al.}(2007)\citenamefont {Falco},
  \citenamefont {Pelster},\ and\ \citenamefont {Graham}}]{Falco07}%
  \BibitemOpen
  \bibfield  {author} {\bibinfo {author} {\bibfnamefont {G.~M.}\ \bibnamefont
  {Falco}}, \bibinfo {author} {\bibfnamefont {A.}~\bibnamefont {Pelster}}, \
  and\ \bibinfo {author} {\bibfnamefont {R.}~\bibnamefont {Graham}},\ }\href
  {\doibase 10.1103/PhysRevA.76.013624} {\bibfield  {journal} {\bibinfo
  {journal} {Phys. Rev. A}\ }\textbf {\bibinfo {volume} {76}},\ \bibinfo
  {pages} {013624} (\bibinfo {year} {2007})}\BibitemShut {NoStop}%
\bibitem [{\citenamefont {Hsueh}\ \emph {et~al.}(2018)\citenamefont {Hsueh},
  \citenamefont {Ong}, \citenamefont {Tseng}, \citenamefont {Tsubota},\ and\
  \citenamefont {Wu}}]{Hsueh18}%
  \BibitemOpen
  \bibfield  {author} {\bibinfo {author} {\bibfnamefont {C.-H.}\ \bibnamefont
  {Hsueh}}, \bibinfo {author} {\bibfnamefont {R.}~\bibnamefont {Ong}}, \bibinfo
  {author} {\bibfnamefont {J.-F.}\ \bibnamefont {Tseng}}, \bibinfo {author}
  {\bibfnamefont {M.}~\bibnamefont {Tsubota}}, \ and\ \bibinfo {author}
  {\bibfnamefont {W.-C.}\ \bibnamefont {Wu}},\ }\href {\doibase
  10.1103/PhysRevA.98.063613} {\bibfield  {journal} {\bibinfo  {journal} {Phys.
  Rev. A}\ }\textbf {\bibinfo {volume} {98}},\ \bibinfo {pages} {063613}
  (\bibinfo {year} {2018})}\BibitemShut {NoStop}%
\bibitem [{\citenamefont {Z\"urn}\ \emph {et~al.}(2013)\citenamefont {Z\"urn},
  \citenamefont {Lompe}, \citenamefont {Wenz}, \citenamefont {Jochim},
  \citenamefont {Julienne},\ and\ \citenamefont {Hutson}}]{Zuern2013}%
  \BibitemOpen
  \bibfield  {author} {\bibinfo {author} {\bibfnamefont {G.}~\bibnamefont
  {Z\"urn}}, \bibinfo {author} {\bibfnamefont {T.}~\bibnamefont {Lompe}},
  \bibinfo {author} {\bibfnamefont {A.~N.}\ \bibnamefont {Wenz}}, \bibinfo
  {author} {\bibfnamefont {S.}~\bibnamefont {Jochim}}, \bibinfo {author}
  {\bibfnamefont {P.~S.}\ \bibnamefont {Julienne}}, \ and\ \bibinfo {author}
  {\bibfnamefont {J.~M.}\ \bibnamefont {Hutson}},\ }\href {\doibase
  10.1103/PhysRevLett.110.135301} {\bibfield  {journal} {\bibinfo  {journal}
  {Phys. Rev. Lett.}\ }\textbf {\bibinfo {volume} {110}},\ \bibinfo {pages}
  {135301} (\bibinfo {year} {2013})}\BibitemShut {NoStop}%
\bibitem [{\citenamefont {G{\"{a}}nger}\ \emph {et~al.}(2018)\citenamefont
  {G{\"{a}}nger}, \citenamefont {Phieler}, \citenamefont {Nagler},\ and\
  \citenamefont {Widera}}]{Gaenger2018}%
  \BibitemOpen
  \bibfield  {author} {\bibinfo {author} {\bibfnamefont {B.}~\bibnamefont
  {G{\"{a}}nger}}, \bibinfo {author} {\bibfnamefont {J.}~\bibnamefont
  {Phieler}}, \bibinfo {author} {\bibfnamefont {B.}~\bibnamefont {Nagler}}, \
  and\ \bibinfo {author} {\bibfnamefont {A.}~\bibnamefont {Widera}},\ }\href
  {\doibase 10.1063/1.5045827} {\bibfield  {journal} {\bibinfo  {journal} {Rev.
  Sci. Instrum.}\ }\textbf {\bibinfo {volume} {89}},\ \bibinfo {pages} {093105}
  (\bibinfo {year} {2018})}\BibitemShut {NoStop}%
\bibitem [{\citenamefont {Grimm}(2007)}]{Grimm2007}%
  \BibitemOpen
  \bibfield  {author} {\bibinfo {author} {\bibfnamefont {R.}~\bibnamefont
  {Grimm}},\ }in\ \href {\doibase 10.3254/978-1-58603-846-5-413} {\emph
  {\bibinfo {booktitle} {{Proceedings of the International School of Physics
  "Enrico Fermi"}}}},\ Vol.\ \bibinfo {volume} {164},\ \bibinfo {editor}
  {edited by\ \bibinfo {editor} {\bibfnamefont {C.~S.}\ \bibnamefont
  {M.~Inguscio}, \bibfnamefont {W.~Ketterle}}}\ (\bibinfo {year} {2007})\ pp.\
  \bibinfo {pages} {413--462}\BibitemShut {NoStop}%
\bibitem [{sup()}]{supps}%
  \BibitemOpen
  \href@noop {} {}\bibinfo {note} {See Supplementary Material}\BibitemShut
  {NoStop}%
\bibitem [{\citenamefont {Naraschewski}\ and\ \citenamefont
  {Stamper-Kurn}(1998)}]{Naraschewski1998}%
  \BibitemOpen
  \bibfield  {author} {\bibinfo {author} {\bibfnamefont {M.}~\bibnamefont
  {Naraschewski}}\ and\ \bibinfo {author} {\bibfnamefont {D.~M.}\ \bibnamefont
  {Stamper-Kurn}},\ }\href {\doibase 10.1103/PhysRevA.58.2423} {\bibfield
  {journal} {\bibinfo  {journal} {Phys. Rev. A}\ }\textbf {\bibinfo {volume}
  {58}},\ \bibinfo {pages} {2423} (\bibinfo {year} {1998})}\BibitemShut
  {NoStop}%
\bibitem [{\citenamefont {Chen}\ \emph {et~al.}(2005)\citenamefont {Chen},
  \citenamefont {Stajic},\ and\ \citenamefont {Levin}}]{Chen2005}%
  \BibitemOpen
  \bibfield  {author} {\bibinfo {author} {\bibfnamefont {Q.}~\bibnamefont
  {Chen}}, \bibinfo {author} {\bibfnamefont {J.}~\bibnamefont {Stajic}}, \ and\
  \bibinfo {author} {\bibfnamefont {K.}~\bibnamefont {Levin}},\ }\href
  {\doibase 10.1103/PhysRevLett.95.260405} {\bibfield  {journal} {\bibinfo
  {journal} {Phys. Rev. Lett.}\ }\textbf {\bibinfo {volume} {95}},\ \bibinfo
  {pages} {260405} (\bibinfo {year} {2005})}\BibitemShut {NoStop}%
\bibitem [{\citenamefont {Kuhn}\ \emph {et~al.}(2007)\citenamefont {Kuhn},
  \citenamefont {Sigwarth}, \citenamefont {Miniatura}, \citenamefont
  {Delande},\ and\ \citenamefont {M{\"{u}}ller}}]{Kuhn2007}%
  \BibitemOpen
  \bibfield  {author} {\bibinfo {author} {\bibfnamefont {R.~C.}\ \bibnamefont
  {Kuhn}}, \bibinfo {author} {\bibfnamefont {O.}~\bibnamefont {Sigwarth}},
  \bibinfo {author} {\bibfnamefont {C.}~\bibnamefont {Miniatura}}, \bibinfo
  {author} {\bibfnamefont {D.}~\bibnamefont {Delande}}, \ and\ \bibinfo
  {author} {\bibfnamefont {C.~A.}\ \bibnamefont {M{\"{u}}ller}},\ }\href
  {\doibase 10.1088/1367-2630/9/6/161} {\bibfield  {journal} {\bibinfo
  {journal} {New J. Phys.}\ }\textbf {\bibinfo {volume} {9}},\ \bibinfo {pages}
  {161} (\bibinfo {year} {2007})}\BibitemShut {NoStop}%
\bibitem [{\citenamefont {Reinaudi}\ \emph {et~al.}(2007)\citenamefont
  {Reinaudi}, \citenamefont {Lahaye}, \citenamefont {Wang},\ and\ \citenamefont
  {Gu\'{e}ry-Odelin}}]{Reinaudi2007}%
  \BibitemOpen
  \bibfield  {author} {\bibinfo {author} {\bibfnamefont {G.}~\bibnamefont
  {Reinaudi}}, \bibinfo {author} {\bibfnamefont {T.}~\bibnamefont {Lahaye}},
  \bibinfo {author} {\bibfnamefont {Z.}~\bibnamefont {Wang}}, \ and\ \bibinfo
  {author} {\bibfnamefont {D.}~\bibnamefont {Gu\'{e}ry-Odelin}},\ }\href
  {\doibase 10.1364/OL.32.003143} {\bibfield  {journal} {\bibinfo  {journal}
  {Opt. Lett.}\ }\textbf {\bibinfo {volume} {32}},\ \bibinfo {pages} {3143}
  (\bibinfo {year} {2007})}\BibitemShut {NoStop}%
\bibitem [{\citenamefont {Vidal}(2004)}]{vidal2004}%
  \BibitemOpen
  \bibfield  {author} {\bibinfo {author} {\bibfnamefont {G.}~\bibnamefont
  {Vidal}},\ }\href {\doibase 10.1103/PhysRevLett.93.040502} {\bibfield
  {journal} {\bibinfo  {journal} {Phys. Rev. Lett.}\ }\textbf {\bibinfo
  {volume} {93}},\ \bibinfo {pages} {040502} (\bibinfo {year}
  {2004})}\BibitemShut {NoStop}%
\bibitem [{\citenamefont {Daley}\ \emph {et~al.}(2004)\citenamefont {Daley},
  \citenamefont {Kollath}, \citenamefont {Schollw\"ock},\ and\ \citenamefont
  {Vidal}}]{daley2004}%
  \BibitemOpen
  \bibfield  {author} {\bibinfo {author} {\bibfnamefont {A.~J.}\ \bibnamefont
  {Daley}}, \bibinfo {author} {\bibfnamefont {C.}~\bibnamefont {Kollath}},
  \bibinfo {author} {\bibfnamefont {U.}~\bibnamefont {Schollw\"ock}}, \ and\
  \bibinfo {author} {\bibfnamefont {G.}~\bibnamefont {Vidal}},\ }\href
  {\doibase 10.1088/1742-5468/2004/04/p04005} {\bibfield  {journal} {\bibinfo
  {journal} {J. Stat. Mech.: Theory Exp.}\ }\textbf {\bibinfo {volume}
  {2004}},\ \bibinfo {pages} {P04005} (\bibinfo {year} {2004})}\BibitemShut
  {NoStop}%
\bibitem [{\citenamefont {White}\ and\ \citenamefont
  {Feiguin}(2004)}]{white2004}%
  \BibitemOpen
  \bibfield  {author} {\bibinfo {author} {\bibfnamefont {S.~R.}\ \bibnamefont
  {White}}\ and\ \bibinfo {author} {\bibfnamefont {A.~E.}\ \bibnamefont
  {Feiguin}},\ }\href {\doibase 10.1103/PhysRevLett.93.076401} {\bibfield
  {journal} {\bibinfo  {journal} {Phys. Rev. Lett.}\ }\textbf {\bibinfo
  {volume} {93}},\ \bibinfo {pages} {076401} (\bibinfo {year}
  {2004})}\BibitemShut {NoStop}%
\bibitem [{\citenamefont {Astrakharchik}\ \emph {et~al.}(2004)\citenamefont
  {Astrakharchik}, \citenamefont {Blume}, \citenamefont {Giorgini},\ and\
  \citenamefont {Pitaevskii}}]{Astra04}%
  \BibitemOpen
  \bibfield  {author} {\bibinfo {author} {\bibfnamefont {G.~E.}\ \bibnamefont
  {Astrakharchik}}, \bibinfo {author} {\bibfnamefont {D.}~\bibnamefont
  {Blume}}, \bibinfo {author} {\bibfnamefont {S.}~\bibnamefont {Giorgini}}, \
  and\ \bibinfo {author} {\bibfnamefont {L.~P.}\ \bibnamefont {Pitaevskii}},\
  }\href {\doibase 10.1103/PhysRevLett.93.050402} {\bibfield  {journal}
  {\bibinfo  {journal} {Phys. Rev. Lett.}\ }\textbf {\bibinfo {volume} {93}},\
  \bibinfo {pages} {050402} (\bibinfo {year} {2004})}\BibitemShut {NoStop}%
\bibitem [{\citenamefont {Tokatly}(2004)}]{tokatly2004}%
  \BibitemOpen
  \bibfield  {author} {\bibinfo {author} {\bibfnamefont {I.~V.}\ \bibnamefont
  {Tokatly}},\ }\href {\doibase 10.1103/PhysRevLett.93.090405} {\bibfield
  {journal} {\bibinfo  {journal} {Phys. Rev. Lett.}\ }\textbf {\bibinfo
  {volume} {93}},\ \bibinfo {pages} {090405} (\bibinfo {year}
  {2004})}\BibitemShut {NoStop}%
\bibitem [{\citenamefont {Fuchs}\ \emph {et~al.}(2004)\citenamefont {Fuchs},
  \citenamefont {Recati},\ and\ \citenamefont {Zwerger}}]{fuchs2004}%
  \BibitemOpen
  \bibfield  {author} {\bibinfo {author} {\bibfnamefont {J.~N.}\ \bibnamefont
  {Fuchs}}, \bibinfo {author} {\bibfnamefont {A.}~\bibnamefont {Recati}}, \
  and\ \bibinfo {author} {\bibfnamefont {W.}~\bibnamefont {Zwerger}},\ }\href
  {\doibase 10.1103/PhysRevLett.93.090408} {\bibfield  {journal} {\bibinfo
  {journal} {Phys. Rev. Lett.}\ }\textbf {\bibinfo {volume} {93}},\ \bibinfo
  {pages} {090408} (\bibinfo {year} {2004})}\BibitemShut {NoStop}%
\bibitem [{\citenamefont {S\"offing}\ \emph {et~al.}(2011)\citenamefont
  {S\"offing}, \citenamefont {Bortz},\ and\ \citenamefont
  {Eggert}}]{soeffing11}%
  \BibitemOpen
  \bibfield  {author} {\bibinfo {author} {\bibfnamefont {S.~A.}\ \bibnamefont
  {S\"offing}}, \bibinfo {author} {\bibfnamefont {M.}~\bibnamefont {Bortz}}, \
  and\ \bibinfo {author} {\bibfnamefont {S.}~\bibnamefont {Eggert}},\ }\href
  {\doibase 10.1103/PhysRevA.84.021602} {\bibfield  {journal} {\bibinfo
  {journal} {Phys. Rev. A}\ }\textbf {\bibinfo {volume} {84}},\ \bibinfo
  {pages} {021602} (\bibinfo {year} {2011})}\BibitemShut {NoStop}%
\bibitem [{\citenamefont {Combescot}\ \emph {et~al.}(2006)\citenamefont
  {Combescot}, \citenamefont {Kagan},\ and\ \citenamefont
  {Stringari}}]{Combescot2006}%
  \BibitemOpen
  \bibfield  {author} {\bibinfo {author} {\bibfnamefont {R.}~\bibnamefont
  {Combescot}}, \bibinfo {author} {\bibfnamefont {M.~Y.}\ \bibnamefont
  {Kagan}}, \ and\ \bibinfo {author} {\bibfnamefont {S.}~\bibnamefont
  {Stringari}},\ }\href {\doibase 10.1103/PhysRevA.74.042717} {\bibfield
  {journal} {\bibinfo  {journal} {Phys. Rev. A}\ }\textbf {\bibinfo {volume}
  {74}},\ \bibinfo {pages} {042717} (\bibinfo {year} {2006})}\BibitemShut
  {NoStop}%
\bibitem [{\citenamefont {Miller}\ \emph {et~al.}(2007)\citenamefont {Miller},
  \citenamefont {Chin}, \citenamefont {Stan}, \citenamefont {Liu},
  \citenamefont {Setiawan}, \citenamefont {Sanner},\ and\ \citenamefont
  {Ketterle}}]{Miller2007}%
  \BibitemOpen
  \bibfield  {author} {\bibinfo {author} {\bibfnamefont {D.~E.}\ \bibnamefont
  {Miller}}, \bibinfo {author} {\bibfnamefont {J.~K.}\ \bibnamefont {Chin}},
  \bibinfo {author} {\bibfnamefont {C.~A.}\ \bibnamefont {Stan}}, \bibinfo
  {author} {\bibfnamefont {Y.}~\bibnamefont {Liu}}, \bibinfo {author}
  {\bibfnamefont {W.}~\bibnamefont {Setiawan}}, \bibinfo {author}
  {\bibfnamefont {C.}~\bibnamefont {Sanner}}, \ and\ \bibinfo {author}
  {\bibfnamefont {W.}~\bibnamefont {Ketterle}},\ }\href {\doibase
  10.1103/PhysRevLett.99.070402} {\bibfield  {journal} {\bibinfo  {journal}
  {Phys. Rev. Lett.}\ }\textbf {\bibinfo {volume} {99}},\ \bibinfo {pages}
  {070402} (\bibinfo {year} {2007})}\BibitemShut {NoStop}%
\bibitem [{\citenamefont {Weimer}\ \emph {et~al.}(2015)\citenamefont {Weimer},
  \citenamefont {Morgener}, \citenamefont {Singh}, \citenamefont {Siegl},
  \citenamefont {Hueck}, \citenamefont {Luick}, \citenamefont {Mathey},\ and\
  \citenamefont {Moritz}}]{Weimer2015}%
  \BibitemOpen
  \bibfield  {author} {\bibinfo {author} {\bibfnamefont {W.}~\bibnamefont
  {Weimer}}, \bibinfo {author} {\bibfnamefont {K.}~\bibnamefont {Morgener}},
  \bibinfo {author} {\bibfnamefont {V.~P.}\ \bibnamefont {Singh}}, \bibinfo
  {author} {\bibfnamefont {J.}~\bibnamefont {Siegl}}, \bibinfo {author}
  {\bibfnamefont {K.}~\bibnamefont {Hueck}}, \bibinfo {author} {\bibfnamefont
  {N.}~\bibnamefont {Luick}}, \bibinfo {author} {\bibfnamefont
  {L.}~\bibnamefont {Mathey}}, \ and\ \bibinfo {author} {\bibfnamefont
  {H.}~\bibnamefont {Moritz}},\ }\href {\doibase
  10.1103/PhysRevLett.114.095301} {\bibfield  {journal} {\bibinfo  {journal}
  {Phys. Rev. Lett.}\ }\textbf {\bibinfo {volume} {114}},\ \bibinfo {pages}
  {095301} (\bibinfo {year} {2015})}\BibitemShut {NoStop}%
\bibitem [{\citenamefont {Orso}(2007)}]{Orso07}%
  \BibitemOpen
  \bibfield  {author} {\bibinfo {author} {\bibfnamefont {G.}~\bibnamefont
  {Orso}},\ }\href {\doibase 10.1103/PhysRevLett.99.250402} {\bibfield
  {journal} {\bibinfo  {journal} {Phys. Rev. Lett.}\ }\textbf {\bibinfo
  {volume} {99}},\ \bibinfo {pages} {250402} (\bibinfo {year}
  {2007})}\BibitemShut {NoStop}%
\bibitem [{\citenamefont {Han}\ and\ \citenamefont {de~Melo}(2010)}]{Han10}%
  \BibitemOpen
  \bibfield  {author} {\bibinfo {author} {\bibfnamefont {L.}~\bibnamefont
  {Han}}\ and\ \bibinfo {author} {\bibfnamefont {C.~S.}\ \bibnamefont
  {de~Melo}},\ }\href {\doibase 10.1088/1367-2630/13/5/055012} {\bibfield
  {journal} {\bibinfo  {journal} {New J.~Phys.~}\ }\textbf {\bibinfo {volume}
  {13}},\ \bibinfo {pages} {055012} (\bibinfo {year} {2010})}\BibitemShut
  {NoStop}%
\bibitem [{\citenamefont {Pilati}\ \emph {et~al.}(2010)\citenamefont {Pilati},
  \citenamefont {Giorgini}, \citenamefont {Modugno},\ and\ \citenamefont
  {Prokof'ev}}]{Pilati2010}%
  \BibitemOpen
  \bibfield  {author} {\bibinfo {author} {\bibfnamefont {S.}~\bibnamefont
  {Pilati}}, \bibinfo {author} {\bibfnamefont {S.}~\bibnamefont {Giorgini}},
  \bibinfo {author} {\bibfnamefont {M.}~\bibnamefont {Modugno}}, \ and\
  \bibinfo {author} {\bibfnamefont {N.}~\bibnamefont {Prokof'ev}},\ }\href
  {\doibase 10.1088/1367-2630/12/7/073003} {\bibfield  {journal} {\bibinfo
  {journal} {New J. Phys.}\ }\textbf {\bibinfo {volume} {12}},\ \bibinfo
  {pages} {073003} (\bibinfo {year} {2010})}\BibitemShut {NoStop}%
\end{thebibliography}%
	
	\appendix*
	\section{Supplementary Material}

\subsection{Experimental parameters}
	\autoref{tab:parameters} lists the experimental parameters for the magnetic fields used in the experiment.
	\begin{table}[b]
		\begin{tabular}{l | r | r | r }
			magnetic field (G) & 763.6 & 832.2 & 900.0 \\
			\hline
			$\Omega_y/2\pi$ (\si{\hertz}) & 22.6 & 23.6 & 24.5 \\
			$a$ ($a_0$) & 4510 & $\infty$ & -7739 \\ 
			interaction parameter & 1.1 & 0.0 & -0.7 \\
			$A$ (\si{\micro\meter}) & 78 & 76 & 72 \\
			$\epot$ (\si{\nano\kelvin}) & 45 & 46 & 44 \\
		\end{tabular}
		\caption{Overview of parameters for different magnetic fields. Scattering lengths taken from \cite{Zuern2013}. The values for $A$ and $\epot$ are averages of all measurement series with different disorder strength.}
		\label{tab:parameters}
	\end{table}
	
	\newpage
	\bibliographystyle{apsrev4-1}
		
\end{document}